\documentclass[12pt]{article}
\usepackage{graphicx}
\begin{document}
\title{Cosmological tachyon condensation}

\author{
Neven Bili\'c$^1$\thanks{bilic@thphys.irb.hr}, Gary B.~Tupper$^2$\thanks{gary.tupper@uct.ac.za}
 and Raoul D.~Viollier$^2$\thanks{raoul.viollier@uct.ac.za}
 \\
$^1$Rudjer Bo\v{s}kovi\'{c} Institute, 10002 Zagreb, Croatia\\
$^2$Centre of Theoretical Physics and Astrophysics,\\
University of Cape Town,  Rondebosch 7701, South Africa\\
}

\maketitle

\begin{abstract}
We consider the prospects for dark matter/energy unification in k-essence type 
cosmologies.
General mappings are established between the k-essence scalar field, the hydrodynamic and braneworld descriptions. We develop an extension of the general relativistic dust model that incorporates the effects of both pressure and the associated acoustic horizon. Applying this to a tachyon model, we show that this inhomogeneous ``variable Chaplygin gas" does evolve into a mixed system containing cold dark matter like gravitational condensate in significant quantities. Our methods can be applied to any dark energy model, as well as to mixtures of dark energy and traditional dark matter.
\end{abstract}

\maketitle



\section{Introduction}

The discovery of the accelerated Hubble expansion in the SNIa data 
\cite{perl1},
combined with observations of the cosmic microwave background
(CMB) 
\cite{hal2,hin3},
 has forced a profound shift in our cosmological paradigm. 
If one makes the conservative assumptions of the validity of Einstein's general relativity and 
the cosmological principle, one concludes that the universe is presently dominated 
by a component that violates the strong energy condition, dubbed {\em dark energy} (DE).
 Moreover, primordial nucleosynthesis constrains the fraction of closure density in baryons,
 $\Omega_{B}$, to a few percent, while galactic rotation curves and cluster dynamics 
imply the existence of a nonbaryonic {\it dark matter} (DM) component 
with $\Omega_{DM} \gg \Omega_{B}$ (for a review see
\cite{peeb4}).
 Currently, the best fit values are $\Omega_{B} = 0.04, \Omega_{DM} = 0.22$ and 
$\Omega_{DE}$ = 0.74 \cite{hin3}. Thus it may be said that we have a firm theoretical 
understanding of only 4\% of our universe.

Pragmatically, the data can be accommodated by 
combining baryons with con\-ven\-tion\-al
cold dark matter (CDM)
candidates and a simple cosmological constant $\Lambda$ providing the DE.
This $\Lambda$CDM model, however, begs the question of why $\Lambda$ 
is non-zero, but 
such that DM and DE are comparable today.
The coincidence problem
of the $\Lambda$CDM model is somewhat ameliorated in
{\it quintessence} models 
 which replace $\Lambda$ by an evolving scalar field.
However,
like its predecessor,
a quintessence-CDM model
assumes that DM and DE are distinct entities.
For a recent review of the most popular DM and DE models,
see \cite{sah5}.

Another interpretation of this data is that
DM/DE are different
manifestations of a common structure. Speculations of this sort were initially made by Hu \cite{hu6}. 
The first definite model of this type was proposed a few years ago 
\cite{kam9,bil7,bil8},
based upon the Chaplygin gas,
a perfect fluid obeying the equation of state
\begin{equation}
p =  - \frac{A}{\rho} \; ,
\label{eq001}
\end{equation}
which has been extensively studied for its
mathematical properties \cite{jack10}.
The general class of models, in which a unification of DM and DE
is achieved through a single entity, is often referred to as
{\em quartessence} \cite{mak11,mak12}.
Among other scenarios of unification that have recently been suggested,
interesting attempts are based on
the so-called {\em k-essence} \cite{chi13,sch14}, a scalar field with
noncanonical kinetic terms which was first introduced as a model for
inflation \cite{arm15}.

The cosmological potential of equation (\ref{eq001})
was first
noted by Kamenshchik {\it et al} \cite{kam9}, who observed that
integrating the energy
conservation equation in a homogeneous model leads to
\begin{equation}
\rho(a) = \sqrt{A + \frac{B}{a^{6}} }  \; ,
\label{eq002}
\end{equation}
where $a$ is the scale factor normalized to
unity today and $B$ an integration constant.
Thus, the Chaplygin gas interpolates between matter,
 $\rho \sim \sqrt{B} a^{-3}$, $p \sim 0$,
at high redshift and a cosmological constant like
$\rho \sim \sqrt{A} \sim - p$ as $a$ tends to infinity.
The essence of the idea in \cite{bil7,bil8}
is simply that
in an {\it inhomogeneous} universe,
highly overdense regions
(galaxies, clusters) have $|w| = |p/\rho| \ll 1$ providing DM, whereas in underdense regions (voids) evolution drives $\rho$ to its limiting value $\sqrt{A}$
giving DE.

Of particular interest is that
the Chaplygin gas 
has an equivalent scalar field formulation
\cite{bil7,bil8,jack10}.
Considering the Lagrangian  
\begin{equation}
{\cal L}  =
 - \sqrt{A} \sqrt{1 - X}\, ,
\label{eq003}
\end{equation}
where
\begin{equation}
X \equiv g^{\mu \nu} \varphi_{, \mu} \varphi_{, \nu} \: ,
\label{eq1203}
\end{equation}
equation (\ref{eq001})
is obtained 
 by evaluating the stress-energy tensor $T_{\mu \nu}$, and  introducing
$u_{\mu} = \varphi_{,\mu} / \sqrt{X}$
for the four-velocity and
$\rho = \sqrt{A} / \sqrt{1 - X}$
for the energy density. One recognizes $\cal{L}$
as a  Lagrangian of the Born-Infeld type, familiar
in the $D$-brane constructions of string/$M$ theory \cite{pol16}.
Geometrically, $\cal{L}$
describes space-time as 
the world-volume of a 
3+1 brane in a 4+1 bulk via
the embedding coordinate $X^{5}$ \cite{sun17}.

To be able to claim that a field theoretical model
actually achieves unification, one must be assured that
initial perturbations can evolve into a deeply nonlinear regime
to form a gravitational condensate of superparticles that
can play the role of CDM. In \cite{bil7, bil8} this was
inferred on the basis of the Zel'dovich
approximation \cite{zel18}. In fact,
for this issue, the 
usual
Zel'dovich approximation has the
shortcoming that the effects of finite sound speed are neglected.

All models that unify DM and DE face the problem of
nonvanishing sound speed and the well-known Jeans instability.
A fluid with a nonzero sound speed has a characteristic scale below which
the pressure effectively opposes gravity. Hence the perturbations of the scale
smaller than the sonic horizon will be prevented from growing.
Soon after the appearance of
\cite{kam9}
and \cite{bil7},
it was pointed out that the perturbative Chaplygin gas (for early work see \cite{fab19},
and more recently \cite{gor20}) 
is incompatible with the observed mass power spectrum \cite{sand21}
and microwave background \cite{cart22}.  
Essentially, these results follow from the adiabatic speed of sound
\begin{equation}
c_s^2 = \left.\frac{\partial p}{\partial \rho}\right|_{s} = \frac{A}{\rho^{2}}
\label{eq004}
\end{equation}
which leads to a comoving acoustic horizon
\begin{equation}
d_s = \int  dt \, \frac{c_s}{a} \; \; .
\label{eq005}
\end{equation}
 The perturbations whose comoving size $R$ is larger than $d_s$ grow as
$\delta = (\rho - \bar{\rho})/\bar{\rho} \sim a$. 
Once the perturbations enter the acoustic horizon, i.e., as soon as $R<d_s$, they undergo damped oscillations.
 In the case of the Chaplygin gas we have
$d_s \sim a^{7/2}/H_0$, where $H_{0}$ is the present day value of the Hubble parameter, reaching Mpc scales already at redshifts of order 10. 
However, to reiterate a point made in \cite{bil7}, 
small perturbations
{\it alone} are not the issue, since large density contrasts are required 
on galactic and cluster scales. 
 As soon as $\delta \simeq 1$ the linear perturbation theory cannot be trusted. 
An essentially nonperturbative approach
is needed in order to investigate whether a significant fraction of initial density perturbations
collapses in gravitationally bound structure - the condensate.
 If that happens the system evolves into 
a two-phase structure - a mixture of
CDM in the form of condensate and DE in the form of uncondensed gas.

The case, where
the Chaplygin gas is mixed with CDM,
has been considered in a
 number of papers \cite{dev23,dev24,avel25,alca26,bean27,amen28,mult29,bert30}.
Here, the Chaplygin gas simply plays the role of DE.
In keeping with the quartessence
philosophy, it would be preferred if  CDM could be replaced by
droplets of Chaplygin gas condensate, as in \cite{bil31}.
Homogeneous world models, containing a mixture of CDM and
Chaplygin gas, have been successfully confronted with lensing statistics
\cite{dev23,dev24}
as well as with supernova and other tests
\cite{avel25,alca26}.

Another 
model, 
the so called ``generalized Chaplygin gas" \cite{ben32},
has gained a wide popularity.
The generalized Chaplygin gas is
defined as \cite{bil7,kam9,ben32}
$p = - A/\rho^{\alpha}$ with $0\leq \alpha\leq 1$ for stability and causality.
As in the Chaplygin gas case, this equation of state 
has an equivalent field theory representation,  
the ``generalized Born-Infeld theory''\cite{ben32,bert33}. However,
the associated Lagrangian
has no equivalent brane interpretation.
The additional parameter does afford greater flexibility: e.g. for small
$\alpha$ the sound horizon is $d_{s} \sim \sqrt{\alpha} a^{2}/H_{0}$,
and thus by fine tuning
$\alpha < 10^{-5}$, 
the data can be perturbatively accommodated \cite{sand21}. 
Bean and Dor\'{e} \cite{bean27} and similarly Amendola {\it et Al}
\cite{amen28} have
examined a mixture of CDM and the generalized Chaplygin
gas against supernova, large-scale structure, and CMB constraints.
They have
demonstrated that a
thorough likelihood analysis favors the limit $\alpha
\rightarrow 0$, i.e. the equivalent to the $\Lambda$CDM
model.  Both papers conclude that the standard Chaplygin gas
is ruled out as a candidate for DE.
However, analysis \cite{bert30,bert33} of the supernova data
seems to indicate that the generalized Chaplygin gas with
$\alpha \geq 1$ is favored over the $\alpha \rightarrow 0$ model and
similar conclusions were drawn in \cite{gor20}. But one should bear in mind that the
generalized Chaplygin gas with $\alpha > 1$ has a superluminal sound speed 
that violates causality  \cite{ell62}.
For a different view on this issue see \cite{bru,kan,bab63}  (see discussion in section
\ref{kessentials}).

The structure formation question, in respect of the Chaplygin gas, was decided in 
\cite{bil34}.
In fact, in the Newtonian approximation, we derived an extension of the spherical model 
\cite{gazt35}
that incorporates nonlinearities in the density contrast $\delta$, as well as the effects of the adiabatic speed of sound. Both are crucial, since for an overdensity 
we have $c_s < \bar{c}_s$,
where $\bar{c}_s$ is the speed of sound of the background.
 Although small initial overdensities follow the expected perturbative evolution,
 for an initial $\delta_R (a_{in})$  exceeding
 a scale $R$-dependent critical $\delta_{c}$,
it was shown that $\delta_R (a)$ tends to infinity at finite redshift, signaling the formation of a bound structure or {\em condensate}. 
Unfortunately, it was further found that, when the required $\delta_c$ is folded with the spectrum of the initial density perturbations to obtain the collapse fraction, less than 1\% of the Chaplygin gas ends up as condensate. Thus the simple Chaplygin gas is not viable due to frustrated structure formation.
In effect, the model is a victim of the
radiation dominated phase, which turns the Harrison-Zel'dovich spectrum
$\delta_{k} \sim k^{1/2}$ to
$\delta_{k} \sim k^{- 3/2}$ at
$R_{\rm CEQ} \simeq  26$ Mpc.
In a pure Chaplygin-gas universe without radiation there would inevitably be sufficient
small scale power to drive condensation.

One way to deal with the structure formation problem, is to assume entropy perturbations 
\cite{hu6,reis36} such that the effective speed of sound vanishes\footnote{ 
Note this ``silent quartessence'' is not different from 
\cite{bil7},
where we tacitly neglected the effects of nonvanishing $c_s$.}.
In that picture we have $\delta p = c_s^2 \delta \rho - \delta A/\rho = 0$
even if $c_s \neq 0$. But as we detail below, in a single field model 
it is precisely the adiabatic speed of sound that governs the evolution. 
Hence, entropy perturbations require
the introduction of a second field on which $A$ depends.
 Aside from negating the simplicity of the one-field model,
some attempts at realizing the nonadiabatic scenario 
\cite{bil37,bil38,bil39}
have convinced us that even if $\delta p = 0$ is arranged as an initial condition, it is all but impossible to maintain this condition in a realistic model for evolution.

The failure of the simple Chaplygin gas does not exhaust all the possibilities for 
quart\-essence.
The Born-Infeld Lagrangian 
(\ref{eq003})
is a special
 case of the string-theory inspired
 tachyon Lagrangian \cite{sen40,gibb41} in which
the constant $\sqrt{A}$ is replaced by a potential $V(\varphi)$
\begin{equation}
{\cal L}  =
 - V(\varphi) \; \sqrt{1 - g^{\mu \nu}
 \, \varphi_{,\mu} \, \varphi_{,\nu} } \; .
\label{eq01}
\end{equation}
In turn, tachyon models are a particular case of k-essence 
\cite{arm15}.
The possibility of obtaining both DM and DE from the tachyon with inverse square potential has been speculated in 
\cite{pad42}.
More recently, it was noted 
\cite{guo43}
that, in a Friedmann-Robertson-Walker (FRW) model, the tachyon model is described by the equation of state (\ref{eq001})
in which the constant $A$
is replaced by a function of the cosmological scale factor $a$, 
so the model was dubbed  ``variable Chaplygin gas''. 
Related models have been examined in \cite{bert44,bert45}, however, those
either produce a larger $d_{s}$  than the simple Chaplygin gas \cite{bert44}, 
or else need fine-tuning \cite{bert45}.\footnote{The tachyon model \cite{bert44}
gives $d_{s} \sim a^{2}/H_{0}$. The two-potential model \cite{bert45} yields 
$d_{s} \sim \sqrt{1-h} a^{2}/H_{0}$,
so it requires $1 - h < 10^{-5}$ like the generalized Chaplygin gas.
Expanding in $1 - h$, the second potential reveals itself to be dominantly a cosmological constant.} 

In this paper we develop a version of the spherical model for studying the evolution of density perturbations even into the fully nonlinear regime. Although similar in spirit to \cite{bil34}, the formalism here is completely relativistic, rather than Newtonian,
and applicable to any k-essence model instead of being restricted to the simple Chaplygin gas. The one key element we carry over from \cite{bil34} is an approximate method for treating the effects of pressure gradients - which is to say the adiabatic speed of sound - on the evolution. Our method is flexible enough and can be extended to deal with a mixture of DE and DM. A spherical model has been applied to DE/DM mixtures in \cite{abra46},
however there the effects of pressure gradients were omitted.

We apply our method to the preliminary analysis of a unifying model based on the tachyon type Lagrangian (\ref{eq01}) with a potential of the form
\begin{equation}
V(\varphi)=V_n \varphi^{2n} \; \; ,
\label{eq08}
\end{equation}
where $n$ is a positive integer. In the regime where structure function takes place, we show that this model effectively
behaves as the variable Chaplygin gas with 
$A(a) \sim a^{6n}$ with $n = 1 (2)$ for a quadratic (quartic) potential. As a result, the much smaller acoustic horizon $d_{s} \sim a^{(7/2+3n)}/H_{0}$ enhances condensate formation by two orders of magnitude over the simple Chaplygin gas ($n = 0$). Hence this type of model
may salvage the quartessence scenario.

The remainder of this paper is organized as follows. In section \ref{kessentials}
 we reformulate k-essence type models in a way that allows us to 
deal with large density inhomogeneities. 
In section 3 we develop the spherical model 
approximation that closes the system of equations. 
These two sections are completely general and stand alone.
Numerical results, in section 4, are presented for positive
power-law potentials and contrasted with the simple Chaplygin gas.
Our conclusions and outlook are given in section \ref{conclude}.
Finally, in \ref{adiabatic}, we derive the adiabatic speed of sound for a general k-essence fluid, and 
in \ref{dbrane}, we give a brief description of
the tachyon model from the braneworld perspective. 

\section{K-essentials}
\label{kessentials}
A minimally coupled k-essence model \cite{arm15,garr47}, is described by
\begin{equation}
S = \int \, d^{4}x \, \sqrt{- g}  \left[ - \frac{R}{16\pi G} + {\cal L} (\varphi,X) \right],
\label{eq4001}
\end{equation}
where ${\cal{L}}$ is the most general Lagrangian, 
which depends on a single  scalar field
$\varphi$ of dimension $m^{-1}$, and on
the dimensionless quantity $X$ defined in (\ref{eq1203}).
For $X>0$ that holds in a cosmological setting, the energy momentum tensor
obtained
from 
(\ref{eq4001})
takes the perfect fluid
form, 
\begin{equation}
T_{\mu \nu}= 2{\cal L}_{X}\:
\varphi_{,\mu}\varphi_{,\nu}
-{\cal L} g_{\mu\nu}=(\rho+p) u_\mu u_\nu - p\, g_{\mu\nu}\, ,
\label{eq03}
\end{equation}
with ${\cal{L}}_{X}$ denoting $\partial {\cal{L}}/\partial X$, and 4-velocity
\begin{equation}
u_\mu = \eta \frac{\varphi_{,\mu}}{\sqrt{X}}\,,
\label{eq511}
\end{equation}
where $\eta$ is $+1$ or  $-1$ according to whether $\varphi_{,0}$ 
is positive or negative, i.e.,
the sign of $u_\mu$
  is chosen so $u_{0}$ is positive.
The associated hydrodynamic quantities  are
\begin{equation}
p ={\cal L}(\varphi,X) ,
\label{eq4003}
\end{equation}
\begin{equation}
\rho = 2 X {\cal L}_{X}(\varphi,X)-{\cal L}(\varphi,X).
\label{eq4004}
\end{equation}

 Two general conditions can be placed upon the functional dependence of ${\cal{L}}$,
\begin{equation}
{\cal{L}}_{X} \geq 0  
\label{eq2.7}
\end{equation}
and
\begin{equation}
{\cal{L}}_{XX} \geq 0 .
\label{eq2.9}
\end{equation}

The first condition (\ref{eq2.7}) stems from the null energy condition,
$T_{\mu \nu} n^{\mu} n^{\nu} \geq 0$ for all light like vectors $n^{\mu}$, 
required for stability \cite{hsu48}.
For a perfect fluid we have  
$T_{\mu \nu}  n^{\mu}  n^{\nu} = ( \rho + p) (u_{\mu} n^{\mu})^{2}$, thus
$\rho + p = 2X {\cal{L}}_{X} \geq 0$, and owing to $X > 0$, we arrive at
(\ref{eq2.7}).

The second condition (\ref{eq2.9}) arises 
from restrictions on the speed of sound.
Observing that (\ref{eq4004}) allows us to view $X$ as a function of $\rho$ and $\varphi$,
 the adiabatic speed of sound 
\begin{equation}
c_s^{2} \equiv \left.\frac{\partial p}{\partial\rho}\right|_{s/n}= \left.\frac{\partial p}{\partial\rho}\right|_\varphi ,
\label{eq4006}
\end{equation}
 as shown in Appendix \ref{adiabatic}, coincides with the so called
{\em effective} speed of sound 
\begin{equation}
c_{s}^{2} = \frac{p_X}{\rho_X} = \frac{{\cal{L}}_{X}}
{{\cal{L}}_{X} + 2 X {\cal{L}}_{XX}}
\label{eq4005}
\end{equation}
obtained in a different way in \cite{garr47}.
 For hydrodynamic stability we require $c_s^2 \geq 0 $.
In addition, it seems physically reasonable to require 
$c_s^2 \leq 1$ in order to avoid possible problems with causality
\cite{ell62}.
Causality violation in relation to superluminal
 sound propagation  
has been the subject of a recent debate \cite{ell62,bru,kan,bab63,bon}.
It has been shown \cite{kan,bon} that if k-essence is to solve
the coincidence problem there must be 
an epoch when perturbations in the k-essence field propagate faster than light.
Hence, on the basis of causality it has been argued \cite{bon} 
that  k-essence models which 
solve the coincidence problem are ruled out as 
physically realistic  candidates for DE. In contrast,
it has been demonstrated \cite{bru,kan,bab63} that 
superluminal sound speed propagation in generic k-essence models does not 
necessarily lead to
causality violation and hence,
the k-essence theories may still be 
legitimate candidates for DE.
As the coincidence problem is not the issue in DE/DM unification models
we stick to $0 \leq c_s^2\leq 1$.
In view of $X \geq 0$ and 
 (\ref{eq2.7})
equation (\ref{eq2.9}) follows.

Formally, one may proceed by solving the $\varphi$ field equation
\begin{equation}
\left( 2 {\cal{L}}_{X} \; g^{\mu \nu} \varphi_{, \nu} \right)_{; \mu} -
{\cal{L}}_{\varphi} = 0,
\label{eq07}
\end{equation}
with ${\cal{L}}_{\varphi}$ denoting $\partial {\cal{L}}/\partial\varphi$,
in conjunction with Einstein's equations to obtain $\rho$ and $p$.
However, it proves more useful to pursue the hydrodynamic picture.
For a perfect fluid 
the conservation equation
 \begin{equation}
{T^{\mu\nu}}_{;\nu}=0
\label{eq000}
\end{equation}
yields, as its longitudinal part $u_\mu T^{\mu\nu}{}_{;\nu} = 0$, the continuity equation
\begin{equation}
 \dot{\rho}+3{\cal H}(\rho+p)=0 ,
\label{eq04}
\end{equation}
and, as its transverse part, the Euler equation
\begin{equation}
 \dot{u}^\mu=\frac{1}{\rho+p}h^{\mu\nu} p_{,\nu}\: ,
\label{eq05}
\end{equation}
where we define
\begin{equation}
 3 {\cal H}  ={u^\nu}_{;\nu}\, ;
\;\;\;\;\;
\dot{\rho} = u^\nu\rho_{,\nu}\, ;
\;\;\;\;\;
{\dot{u}^\mu}=u^\nu {u^\mu}_{;\nu}\,.
\label{eq101}
\end{equation}
The tensor
\begin{equation}
 h_{\mu\nu}=g_{\mu\nu}-u_\mu u_\nu 
\label{eq06}
\end{equation}
is a projector onto the three-space orthogonal to $u^\mu$.
The quantity ${\cal H}$ is the local Hubble parameter.
Overdots indicate the proper time derivative.

Now, using (\ref{eq511})-(\ref{eq4004}) the $\varphi$ field equation (\ref{eq07})
can be expressed as
\begin{equation}
\left(\eta\, \frac{\rho + p}{\sqrt{X}} u^{\mu} \right)_{; \mu} - {\cal{L}}_{\varphi} = 0,
\label{eq024}
\end{equation}
or
\begin{equation}
\dot{\rho} + 3 {\cal{H}} ( \rho + p) + ( \dot{\varphi} -\eta \sqrt{X} ) {\cal{L}}_{\varphi}=0.
\label{eq025}
\end{equation}
 where, as before, $\eta={\rm sgn}\, (\varphi_{,0})$.
Hence, for ${\cal{L}}_{\varphi} = 0$ (purely kinetic k-essence), the $\varphi$ field equation 
(\ref{eq07}) is equivalent to (\ref{eq04}).
In the general case, i.e. for ${\cal{L}}_{\varphi}\neq 0$, 
equation (\ref{eq07})  together with (\ref{eq04}) implies 
\begin{equation}
{\dot{\varphi}}^2 = X( \varphi, \rho) \, ,
\label{eq026}
\end{equation}
provided (\ref{eq4004}) is invertible.

Next, we observe that Euler's equation can be written in various forms. 
Equation
\begin{equation}
\dot{u}_{\mu} = \frac{h_{\mu}^{\nu}  X_{,\nu}}{2 X}
\label{eq027}
\end{equation}
follows directly from (\ref{eq511})-(\ref{eq4004}) and (\ref{eq101}). 
With $X$ a function of $\varphi$ and $\rho$, 
the pressure $p$ also becomes a function of $\varphi$ and $\rho$. Thus 
by (\ref{eq511}), (\ref{eq4006}) and (\ref{eq06})
we find
\begin{equation}
\dot{u}_{\mu} = \frac{c_{s}^{2}}{\rho + p}  h_{\mu}^{\nu} \rho_{, \nu} \, .
\label{eq028}
\end{equation}
This is a simple demonstration of the observation made earlier, that in a single component system it is the adiabatic (rather than effective) speed of sound that controls evolution.\footnote{It seems to us that this point is often confused in the literature.}
Rather than specifying ${\cal{L}}$ directly, one may choose $p= p( \varphi, \rho)$ and then find $c_{s}^{2}$ from (\ref{eq4003}). Up to an overall multiplicative integration function of $\varphi$ only (which in turn can be absorbed in a reparameterization
of $\varphi$ itself) equations (\ref{eq027}) and (\ref{eq028}) imply
\begin{equation}
X(\varphi, \rho) = {\rm exp} \left(2
\displaystyle{\int \; \frac{c_{s}^{2} \; d \rho}{\rho + p}} \right) \; \; ,
\label{eq029}
\end{equation}
which may be used in (\ref{eq026}) as an evolution equation for $\varphi$, while (\ref{eq04}) is an evolution equation for $\rho$.
 Further, equation (\ref{eq029}) can be formally inverted to 
give $\rho$ as a function of $\varphi$ and $X$, 
thus allowing the construction of
${\cal{L}} = p \left( \varphi, \rho (\varphi, X) \right)$. 
An example of this will be given in section 4. However, first we need to close the system of evolution equations.
\section{The Spherical Model}

\setcounter{equation}{29}
Since the 4-velocity 
(\ref{eq511}) 
is derived from a potential, the associated rotation tensor vanishes identically. The Raychaudhuri equation for the velocity congruence assumes a simple form
\begin{equation}
3 {\dot{\cal{H}}} + 3 {{\cal{H}}^{2}} + \sigma_{\mu\nu}\sigma^{\mu\nu}+ u^{\mu} u^{\nu} \; R_{\mu \nu} = 
\dot{u}^\mu{}_{;\mu} \, .
\label{eq030}
\end{equation}
 with the shear tensor defined as
\begin{equation}
\sigma_{\mu\nu}=
{h^\alpha}_\mu {h^\beta}_\nu u_{(\alpha;\beta)} - {\cal H} h_{\mu\nu}\: .
\label{eq103}
\end{equation}
We thus obtain an evolution equation for ${\cal{H}}$ that appears in 
(\ref{eq04}),
 sourced by gravity through the Ricci tensor 
${\cal{R}}_{\mu \nu}$ and by 
both shear and
the divergence of the acceleration $\dot{u}^{\mu}$. 
If $\dot{u}^{\mu} = 0$, as for dust, equations 
(\ref{eq04})
 and (\ref{eq030}), together with
Einstein's equations for $R_{\mu \nu}$ comprise the spherical model \cite{gazt35}. 
However, we are not
interested in dust, since generally $\dot{u}^{\mu} \neq 0$ as given by Euler's equation 
(\ref{eq028}).
Indeed, this term is responsible for the Jeans phenomenon in perturbation theory. One is only allowed to neglect $\dot{u}^{\mu}$ in the long wavelength limit, where everything clusters, but one has no realistic information about the small (i.e. subhorizon) scales.

The spherical top-hat profile is often invoked to justify 
neglecting the acceleration term (see e.g. \cite{abra46} and references therein).
In fact, this leads to infinite pressure forces on the bubble boundary. Even if suitably regularized, the influence of these large forces on the bubble evolution is never accounted for. Needless to say, this makes the reliability of the inferences highly problematic, unless one invokes entropy perturbations again.

For a one-component model, the Raychaudhury equation  
(\ref{eq030}) combines with Einstein's equations to
\begin{equation}
3\dot{\cal H}+3{\cal H}^2
+\sigma^2
+4\pi G (\rho +3p)=
\left(\frac{c_s^2 h^{\mu\nu} \rho_{,\nu}}{p+\rho}\right)_{;\mu}\, .
\label{eq008}
\end{equation}
with
\begin{equation}
\sigma^2=\sigma_{\mu\nu}\sigma^{\mu\nu}.
\label{eq102}
\end{equation}

 In general, the 4-velocity $u^{\mu}$ can be decomposed as \cite{bil49}
\begin{equation}
u^{\mu} = \left( U^{\mu} + v^{\mu} \right) / \sqrt{1 - v^{2}} \; \; ,
\label{eq031}
\end{equation}
where $U^{\mu} = \delta_{0}^{\mu} / \sqrt{g_{00}}$ is the 4-velocity  of fiducial observers at rest in the coordinate system, and $v^{\mu}$ is spacelike, with $v^{\mu} v_{\mu} = - v^{2}$ and $U^{\mu} v_{\mu} = 0$.
In a multi-component model (e.g. dark energy plus CDM, 
or a mixture of condensed and uncondensed k-essence)
one can, in the first approximation, neglect the relative
peculiar velocities. Then there
 will be a pair of equations (\ref{eq04})
and (\ref{eq008}) 
for each component with $\rho + 3 p$ replaced by a sum over all components
and  each component may be treated in comoving coordinates.
In comoving coordinates  $v^\mu=0$ and
\begin{equation}
 u_\mu=\sqrt{g_{00}}\delta^0_\mu;
\hspace{.5cm}
 u^\mu=\frac{1}{\sqrt{g_{00}}}\delta_0^\mu;
 \hspace{.5cm}
 h_{00}=h_{0i}=0;
\hspace{.5cm} 
h_{ij}=g_{ij}.
\label{eq061}
\end{equation}
Then the nonvanishing components
of the shear tensor are $\sigma_{ij}$ and from (\ref{eq103})
it follows
\begin{equation}
\sigma_{ij}=
u_{(i;j)} - {\cal H} h_{ij}\: ,
\label{eq105}
\end{equation}
Assuming
\begin{equation}
ds^2 =N^2dt^2 - \gamma_{ij} dx_i dx_j
\label{eq107}
\end{equation}
we find
\begin{equation}
\sigma_{ij}= -\frac{1}{2N} \partial_t \gamma_{ij}
+ {\cal H} \gamma_{ij}\: .
\label{eq106}
\end{equation}
In spherically symmetric spacetime it is convenient to write the metric in the form
\begin{equation}
ds^2 =N(t,r)^2dt^2 - b(t,r)^2(dr^2+r^2f(t,r)d\Omega^2) 
\label{eq09}
\end{equation}
where $N(t,r)$ is the lapse function, $b(t,r)$ is the local expansion scale,
and $f(t,r)$ describes the departure from the flat space for which $f=1$.
We assume that $N$, $a$, and $f$  are arbitrary functions of $t$ and $r$ which are
regular
and different from zero  at $r=0$.
Then, the local Hubble paprameter and the shear are given by
\begin{equation}
{\cal H}=\frac{1}{N}\left(\frac{b_{,0}}{b}+
\frac{1}{3}\frac{f_{,0}}{f}\right)
\label{eq160}
\end{equation}
and
\begin{equation}
\sigma^2=\sum_i\sigma^{ii}\sigma_{ii}
=\frac{2}{3}\left(\frac{1}{2N} \frac{f_{,0}}{f}\right)^2 .
\label{eq111}
\end{equation}

In addition to the spherical symmetry we also require 
an FRW spatially flat asymptotic geometry, i.e., 
for $r\rightarrow \infty$ we demand 
 \begin{equation}
N\rightarrow  1; \hspace{.5cm} f\rightarrow 1;
\hspace{.5cm}
b\rightarrow {a}(t).
\label{eq091}
\end{equation}
Here $a$ denotes the usual expansion scale.

The righthand side of (\ref{eq008}) is difficult to treat in full generality.
As in \cite{bil34}, we apply the ``local approximation" to it: The density contrast $\delta = (\rho - \bar{\rho})/\bar{\rho}$ is assumed to be of fixed Gaussian shape with comoving size $R$, but time-dependent amplitude, so that
\begin{equation}
    \rho (t,r) = \bar{\rho}(t)[1+ \delta_{R}(t) 
  \, 
  e^{-r^2/(2 R^2)}].
\label{eq308}
\end{equation}
and the spatial derivatives are evaluated at the origin. This is in keeping with the spirit of the spherical model, where each region is treated as independent.

Since  $\partial_i \rho=0$ at $r=0$,
naturally $\partial_i N=0$ 
and $\partial_i b=0$ at $r=0$.
Hence,
\begin{equation}
 N(t,r)= N(t,0)(1+{\cal{O}} (r^2)); \hspace{1cm}
 b(t,r)=b(t,0)(1+{\cal{O}} (r^2)).
\end{equation}
Besides,  one finds $f_{,0}\rightarrow 0$ as  $r\rightarrow 0$ which  follows from
 Einstein's equation ${G^1}_0=0$
\begin{equation}
 2\frac{b_{,01}}{b} +\frac{b_{,1}}{b}
\left( \frac{f_{,0}}{f}-2\frac{b_{,0}}{b}\right)
-\frac{N_{,1}}{N}\left( 2\frac{b_{,0}}{b}+ \frac{f_{,0}}{f}\right)+
\frac{f_{,0}}{f} \left(\frac{1}{r}-\frac{1}{2}\frac{f_{,1}}{f} \right)=0.
\end{equation}
Since the first three terms on the lefthand side vanish as $r\rightarrow 0$
the last term can vanish if and  only if $f_{,0}/f=0$ at $r=0$.
By (\ref{eq111}) the shear scalar $\sigma$ vanishes at the origin.

From now on we denote by ${\cal H}$, $b$, and $N$ the correspobding functions
of $t$ and $r$ evaluated at $r=0$, i.e.,
 ${\cal H}\equiv {\cal H}(t,0)$,
$b\equiv b(t,0)$ and $N\equiv N(t,0)$.
 According to (\ref{eq160}), the local  Hubble parameter at the origin is related to the
local expansion scale as 
\begin{equation}
{\cal H}=\frac{1}{N b}\frac{db}{dt}
\label{eq0401}
\end{equation}
Evaluating (\ref{eq008}) at the origin
yields our working approximation to
the Raychaudhuri equation, i.e.
 we obtain 
\begin{equation}
\frac{1}{N}\frac{d\cal H}{dt}+{\cal H}^2
+\frac{4\pi G}{3} (\rho +3p)=
\frac{c_s^2(\rho-\bar{\rho})}{b^2R^2(p+\rho)}\, .
\label{eq040}
\end{equation}
The recommendations of this equation are that it extends the spherical dust model, by incorporating both pressure and, via the speed of sound, the Jeans' phenomenon. In particular, it reproduces the linear theory with the identification $k = \sqrt{3}/R$ for the wavenumber.
\section{Cosmological Tachyon Condensation}
We will now apply our formalism to a particular subclass of k-essence unification 
models described by  (\ref{eq01}).
However, first it is useful to see how such models can be reconstructed using the methods of section 2.

Violating the strong energy condition with positive $\rho$ requires $p < 0$, while stability demands $c_{s}^{2} = \partial p/ \partial p \geq 0$. These criteria are met by\footnote{We do not consider the trivial generalization of adding a function of $\varphi$ alone to $p$.}
\begin{equation}
p = - \frac{A ( \varphi)}{\rho^{\alpha}}\, , \;\;\; \;\;\;
A (\varphi) > 0,
\label{eq045}
\end{equation}
for which
\begin{equation}
c_{s}^{2} = \frac{\alpha  A ( \varphi)}{\rho^{\alpha + 1}} \geq 0 , \;\;\;  \;\;\;
\alpha > 0  .
\label{eq046}
\end{equation}
Note that when the null energy condition is saturated, we have $\rho^{\alpha + 1} = A(\varphi)$, and that
causality restricts $\alpha$ to  $\alpha \leq 1$. Using (\ref{eq029}),
we arrive at
\begin{equation}
X (\varphi, \rho) = \Bigg[ 1 - \frac{A ( \varphi)}{\rho^{1 + \alpha}} 
 \Bigg]^{2 \alpha/(1 + \alpha)}
\label{eq047}
\end{equation}
and the Lagrangian density of the scalar field
\begin{equation}
{\cal{L}} = -  A ( \varphi)^{\alpha/(1 + \alpha)} 
\left[ 1 - X^{(1 + \alpha)/2 \alpha} \right]^{1/(1 + \alpha)}  .
\label{eq048}
\end{equation}
Only for $\alpha = 1$ does one have $c_{s}^{2} = 1$ at the point where the null energy condition is saturated.
Moreover, only for $\alpha = 1$ can one obtain the tachyon Lagrange density
\begin{equation}
{\cal{L}} = - \sqrt{A(\varphi)}  \sqrt{1 - X} \, ,
\label{eq049}
\end{equation}
which coincides with (\ref{eq01}), identifying
$A (\varphi) = V( \varphi )^2$.  The equation of state is then given by 
\begin{equation}
 p = 
- \frac{V(\varphi)^2}{\rho}\, ,
\label{eq049a}
\end{equation}
and the quantity $X$ may be expressed as
\begin{equation}
X(\rho, \varphi) = 1-\frac{V(\varphi)^2}{\rho^2} = 1 - c_{s}^{2} = 1 + w \; \; .
\label{eq050}
\end{equation}
Finally, only for $\alpha = 1$, can the tachyon model be reinterpreted as a 3 + 1
brane, moving in a warped 4 + 1 spacetime, as shown in Appendix B.

Equations ,
(\ref{eq04}),
(\ref{eq026}),
(\ref{eq0401}), and (\ref{eq040}) determine the
evolution of the density contrast. However, as this set of equation is not complete,
it must be supplemented by a similar set of equations for the background
quantities  $\bar{\rho}$ and $H$
\begin{equation}
 \frac{d \bar{\rho}}{dt} + 3(\bar{\rho}+\bar{p})=0 ,
\label{eq17}
\end{equation} 
\begin{equation}
\frac{d H}{dt}+H^2
+\frac{4\pi G}{3} (\bar{\rho} +3\bar{p})=0,
\label{eq18}
\end{equation}
where $\bar{p}=p(\bar{\rho},\bar{\varphi})$. 
Due to (11)
the field $\varphi$ in comoving coordinates is a function of time only,
 its gradient is always timelike or null, i.e., $X\geq 0$,
 and the perfect fluid description
remains valid even in a deep non-linear regime.
In comoving coordinates equation (26)
 reads
\begin{equation}
\left(\frac{d\varphi}{dt}\right)^2 = N^2 X (\varphi, \rho).
\label{eq1}
\end{equation}
In particular,
in the asymptotic region $r\rightarrow \infty$ we have
\begin{equation}
\left(\frac{d\varphi}{dt}\right)^2 = X (\varphi,\bar{\rho}).
\label{eq2}
\end{equation}
Equating (\ref{eq1}) with (\ref{eq2}) we find
an expression for the local lapse function $N$ in terms of $X$
\begin{equation}
N \equiv \sqrt{ X (\varphi, \bar{\rho}) / X (\varphi, \rho )}\, .
\label{eq6}
\end{equation}
Hence, the complete set of equations for $\bar{\rho}$, $H$, $\varphi$, 
$b$, $\rho$, and $\cal H$, consists of (\ref{eq17}), (\ref{eq18}),
(\ref{eq2}), and
\begin{equation}
\frac{db}{dt}=N b{\cal H} ,
\label{eq161}
\end{equation}
\begin{equation}
\frac{d\rho}{dt} + 3 N \; {\cal{H}} \; (\rho + p) = 0,
\label{eq4}
\end{equation}
\begin{equation}
\frac{d\cal{H}}{dt} + N \; \left[ {\cal{H}}^{2} +
\frac{4 \pi G}{3} \; (\rho + 3 P) -
\frac{c_{s}^{2} \; (\rho - \bar{\rho})}{b^{2} R^{2} (\rho + p)} \right] = 0,
\label{eq5}
\end{equation}
where $N$ is given by (\ref{eq6}).
In this way, we have a system of six coupled ordinary differential equations
that describes the evolution of both the background and the spherical inhomogeneity.
The definition of the Hubble parameter
\begin{equation}
H=\frac{1}{a}\frac{da}{dt}
\label{eq20}
\end{equation}
is used
to express the evolution in terms of the background scale factor $a$.

Here we restrict our attention to the power-law potential (\ref{eq08}). In the high density regime, where $c_{s}^{2}$ is small, we have $X \simeq 1$, and (\ref{eq026}) can be integrated yielding 
$\varphi \simeq 2/(3 H)$, where $H \simeq H_{0} \sqrt{\Omega} a^{-3/2}$, $\Omega$ being the equivalent matter content at high redshift. Hence, as promised, $A(\varphi)=V(\varphi)^2 \sim a^{6n}$, which leads to a suppression of 10$^{-6}$ 
at $z = 9$ for $n = 1$.

 To proceed we require a value for the constant $V_{n}$ in the potential (\ref{eq08}). 
Changing $V_{n}$ effects not only the pressure today but also the speed of sound and thereby the amount of structure formation. 
One must also be mindful that if there is a large amount of nonlinear structure formation then the single fluid description will break down at low redshift.
 As the main purpose of this paper is to investigate the evolution of inhomogeneities
we will not pursue the exact fitting of the background evolution.
 Hence, rather than attempting to fit $V_{n}$ using the  naive background model, 
we estimate  $V_{n}$ as follows. We integrate (\ref{eq026}) with 
\begin{equation}
X=1+w(a)\simeq 1-\frac{\Omega_\Lambda}{\Omega_\Lambda+\Omega a^{-3}}\, ,
\hspace{1cm}
 \Omega + \Omega_{\Lambda} = 1,
\end{equation}
as in a $\Lambda$CDM universe \cite{bert44} and obtain
\begin{equation}
\varphi (a) \simeq \frac{2}{3 H_{0}\sqrt{\Omega_{\Lambda}}}  \, \arctan \left(
\sqrt{ \frac{\Omega_{\Lambda}}{\Omega}} \, a^{3/2} \right).
\label{eq055}
\end{equation}
We then fix the pressure given by (\ref{eq01}) to equal that of $\Lambda$ at $a = 1$, i.e.
\begin{equation}
- \rho_{0} \; \Omega_{\Lambda} = - V_n\, \varphi(1)^{2n} \; \sqrt{\Omega_{\Lambda}} \; \; ,
\label{eq056}
\end{equation}
yielding  
\begin{equation}
V_{n} = \frac{3 \alpha_{n}}{8 \pi G} \; H_{0}^{2(n+1)} \; \; ,
\label{eq057}
\end{equation}
where
\begin{equation}
\alpha_n = \sqrt{\Omega_{\Lambda}} \; \left[
\frac{2}{3 \sqrt{\Omega_{\Lambda}}} \; \arctan \left(
\sqrt{ \frac{\Omega_{\Lambda}}{\Omega}} \right) \right]^{- 2n} ,
\label{eq058}
\end{equation}
so $\alpha_{0} \simeq 0.854$, $\alpha_{1} \simeq 1.34$,  and $\alpha_{2} \simeq 2.09$.
With these values
the naive background in our model reproduces the standard cosmology 
from decoupling up to the scales of about $a=0.8$ and fits the cosmology today only approximately.

Using (\ref{eq20}), we solve our differential equations with
$a$ starting from the initial $a_{\rm dec}=1/(z_{\rm dec}+1)$  
at decoupling  redshift $z_{\rm dec}=1089$ for a particular
comoving size $R$. 
The initial conditions for the background are given by
\begin{equation}
\bar{\rho}_{\rm in}=\rho_0\frac{\Omega}{a_{\rm dec}^3};
 \hspace{1cm}
H_{\rm in}=H_0\sqrt{\frac{\Omega}{a_{\rm dec}^3}};
\hspace{1cm}
\bar{\varphi}_{\rm in} = \frac{2}{3 H_{\rm in}},
\label{eq23}
\end{equation}
and for the initial inhomogeneity we take
\begin{equation}
\rho_{\rm in}=\bar{\rho}_{\rm in} (1+\delta_{\rm in}) \; , 
 \hspace{0.75cm}
{\cal H}_{\rm in}=H_{\rm in}\left(1-\frac{\delta_{\rm in}}{3}\right) \; ,
\hspace{0.75cm}
\varphi_{\rm in}=\bar{\varphi}_{\rm in}= \frac{2}{3 H_{\rm in}},
\label{eq25}
\end{equation}
where $\Omega=0.27$ represents the effective dark matter fraction
and $\delta_{\rm in}=\delta_R (a_{\rm dec})$ is a variable initial density
contrast, chosen arbitrarily for a particular $R$.

Here it is worthwhile mentioning that 
the set of equations
(\ref{eq17}), (\ref{eq18}),
(\ref{eq2}), and
(\ref{eq161})-(\ref{eq5}) preserves
 the condition $X\geq 0$ and hence,
the perfect fluid description  remains valid
throughout
the nonlinear evolution described above.
The evolution starts from a small initial value of $\varphi_{\rm in}$ and large
$\rho_{\rm in}$ so that initially $X\simeq 1$ and we take initial $\dot{\varphi}$
to be positive.
According to (\ref{eq050}), as $\varphi$ increases and $\rho$ decreases,  
$X$ decreases up to a point where it becomes 0.
The first such point may be roughly at $a$ between 0.1 and 1.
At that point
the sign of  $\dot{\varphi}$  flips and $X$
remains positive up to the next point at which $X=0$.
The sign of  $\dot{\varphi}$  flips again keeping $X$ positive, and so on.
Basically, this process continues {\it ad infinitum} never violating $X\geq 0$.
Hence, the perfect fluid assumption is legitimate
even in the deep non-linear regime.

\begin{figure}
\begin{center}
\includegraphics[width=.5\textwidth,trim= 0 2cm 0 2cm]{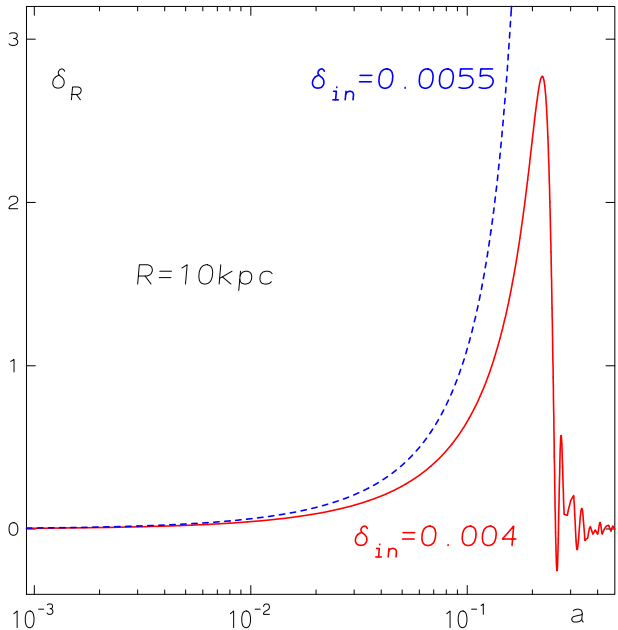}
\caption{
 Evolution of $\delta_{R}(a)$ in
            the tachyon spherical model
           from $a_{\rm dec} = 1/1090$
           for $n=2$, $R$ = 10 kpc,
           $\delta_{R} (a_{\rm dec})$ =0.004 (solid)
            and $\delta_{R} (a_{\rm dec})$ =0.0055 (dashed).
}\label{fig1}
\end{center}
\end{figure}

\begin{figure}
\begin{center}
\includegraphics[width=.5\textwidth,trim= 0 2cm 0 2cm]{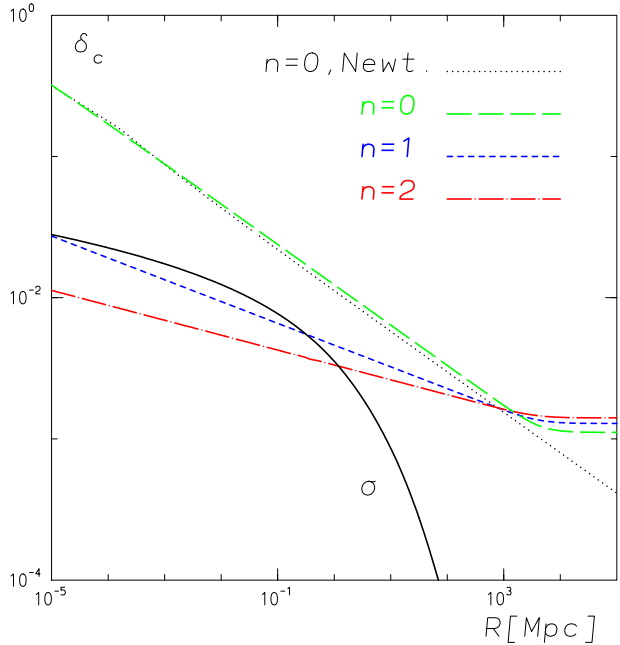}
\caption{
Initial value $\delta_{R}(a_{\rm dec})$  versus $R$
for $\Omega=0.27$ and $h=0.71$.
The threshold $\delta_c (R)$
is shown by the line separating
 the condensation regime from the damped oscillations regime.
The solid line gives $\sigma (R)$ calculated using the
concordance model.
 }\label{fig2}
\end{center}
\end{figure}

In figure \ref{fig1} the representative case of evolution of two initial
perturbations starting from 
decoupling
for $R$ = 10 kpc is shown  for $n = 2$.
 The plots  represent two distinct regimes: 
the growing mode or condensation (blue dashed line) and the damped oscillations
(red solid line).
In contrast to the linear theory, where for any $R$ the acoustic
horizon will eventually stop $\delta_R$ from growing, irrespective of the initial
value of the perturbation, here we have for an
initial $\delta_{R} (a_{\rm dec})$ above a certain
threshold $\delta_{c} (R)$,
$\delta_{R} (a) \rightarrow \infty$ at
finite $a$, just as in the dust model.
 Thus perturbations with $\delta_{R} (a_{dec}) \geq \delta_{c} (R)$ evolve into
a {\it nonlinear} gravitational condensate that at low $z$ behaves as pressureless super-particles.
Conversely, for a sufficiently small
$\delta_{R} (a_{\rm dec})$, the acoustic horizon can stop
$\delta_{R} (a)$ from growing;
 at low redshift the perturbations behave as expected from {\it linear} theory.
Figure \ref{fig2} shows how the threshold
$\delta_{c}(R)$ divides the two regimes
depending on the comoving scale $R$.

The crucial question now  is what fraction of the tachyon gas goes into condensate.
In \cite{bil31} it was shown that if this fraction was sufficiently large, the
CMB and the mass power spectrum
could be reproduced for the simple Chaplygin gas.
To answer this question quantitatively, 
we follow the  Press-Schechter
procedure \cite{pres53} as in  \cite{bil34}.
Assuming $\delta_{R}(a_{\rm dec})$ is given by a Gaussian random field with
dispersion $\sigma(R)$, and
including the notorious factor of 2, to
account for the cloud in cloud problem,
the condensate fraction at
a scale $R$ is given by
\begin{equation}
F (R) = 2 \int_{\delta_{c}(R)}^{\infty} \; \frac{d \delta}{ \sqrt{2 \pi} \sigma (R) }\;
{\rm exp} \left( - \frac{\delta^{2}}{2 \sigma^{2}(R)} \right)
= {\rm erfc} \left(
\frac{ \delta_{c} (R) }{ \sqrt{2} \; \sigma (R) }
\right) \, ,
\label{eq401}
\end{equation}
where $\delta_c(R)$ is the threshold 
shown in figure \ref{fig2}. In figure \ref{fig2} we also exhibit
the dispersion
\begin{equation}
\sigma^{2}(R) = \int_{0}^{\infty} \; \frac{dk}{k} \; {\rm exp}
( - k^{2} R^{2} ) \Delta^{2} (k, a_{\rm dec}) ,
\label{eq402}
\end{equation}
calculated using the Gaussian window function  and the 
variance of the concordance
model
\cite{hin3} 
\begin{equation}
\Delta^{2} (k,a)={\rm const} \left(\frac{k}{a H}\right)^4
T^2(k)\left(\frac{k}{7.5 a_0 H_0}\right)^{n_{s}-1}\, .
\label{eq403}
\end{equation}
In figure 3 we present $F(R)$, calculated
using (\ref{eq401})-(\ref{eq403}) with
const=7.11$\times 10^{-9}$, the spectral index $n_{s}$=1.02,
and the parameterization of Bardeen {\it et al} \cite{bard54} for the transfer
function $T(k)$ with $\Omega_{B}$=0.04.
The parameters are fixed by fitting (\ref{eq403})
to the 2dFGRS power spectrum data \cite{perc55}.
Our result
demonstrates that the collapse fraction is about
70\% for $n=2$ for a wide range of the comoving size $R$
 and peaks at about 45\% for $n = 1$.

Albeit encouraging, these preliminary results do not in themselves demonstrate that 
the tachyon with potential (8) constitutes a  viable cosmology. Such a step requires the inclusion of baryons and comparison with the full cosmological data, much of which obtains at low redshift. What has been shown is that it is not valid in an adiabatic model to simply pursue linear perturbations to the original background : the system evolves nonlinearly into a mixed system of gravitational condensate and residual k-essence so that the ``background'' at low $z$ is quite different from the initial one. 
Because of this one needs new computational tools for a meaningful confrontation with the data. 

\begin{figure}
\begin{center}
\includegraphics[width=.5\textwidth,trim= 0 2cm 0 2cm]{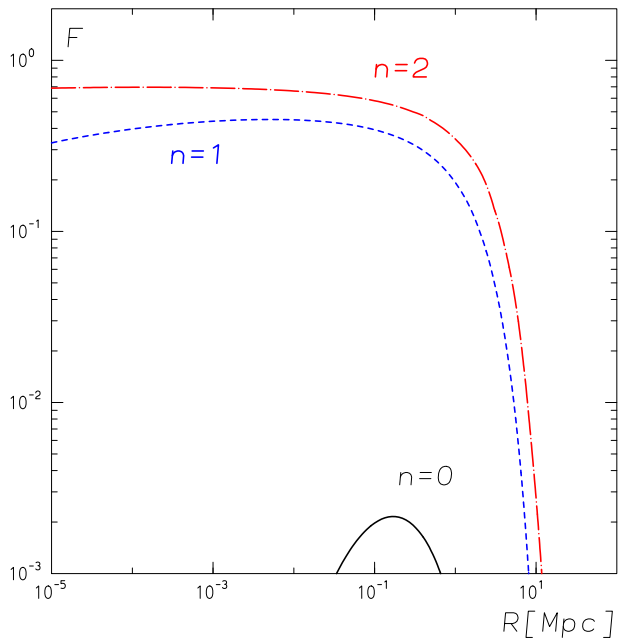}
\caption{
Fraction of the tachyon gas in collapsed objects using
$\delta_c(R)$ and $\sigma(R)$ from
figure \ref{fig2}.}
\label{fig3}
\end{center}
\end{figure}

\section{Summary and Conclusions}
\label{conclude}
The first key test for any proposed quartessence model should be: Does it actually yield nonlinear dark matter structure, as well as linear dark energy in the {\it inhomogeneous} 
{\it almost} FRW universe that we see. In this paper
we have analyzed the nonlinear evolution of the  tachyon-like k-essence
with a very simple potential $V=V_0\varphi^{2n}$.
We have demonstrated that a significant fraction of the fluid, in particular for $n=2$,
collapses into condensate objects that play the role of cold dark matter.
No dimensionless fine tunings were required
beyond the inevitable $\sqrt{G} H_{0} \ll 1$.

Moreover, these results were obtained in a relativistic framework for nonlinear evolution that is as simple as the spherical dust model but includes the key effects of the acoustic horizon. Although it could be subject to possible improvements (e.g. a variable 
Gaussian width \cite{bil34}, and it is lacking the sophistication of the exact spherical model, 
\cite{sus56}, 
it does allow us to make the sort of quantitative
assessments that have been missing \cite{avel57}.
It is directly formulated in a convenient coordinate gauge, does not involve any hidden assumptions, and it easily deals with multi-component systems.

The  tachyon k-essence unification remains to be tested against large-scale structure and CMB observations. However, we maintain, contrary to the opinion advocated in \cite{sand21},
that the sound speed problem may be alleviated in unified models no more unnatural 
than the $\Lambda$CDM model. Indeed, an encouraging feature of the positive power-law potential is that it provides for acceleration as a periodic transient phenomenon 
\cite{fro58}
which obviates the de Sitter horizon problem \cite{bil37} , and,
we speculate, could even be linked to inflation. 

\appendix

\section{Adiabatic speed of sound}
\label{adiabatic}
The standard definition of the adiabatic speed of sound is
\begin{equation}
 c_s^2=\left.\frac{\partial p}{\partial\rho}\right|_{s/n} \; \; ,
\label{eq2601}
\end{equation}
where the differentiation is taken  at constant $s/n$,
i.e. for an isentropic process.
Here  $s=S/V$ is the entropy density and $n=N/V$  the particle number density
associated with the particle number $N$. 
We use the terminology and notation of Landau and Lifshitz \cite{land59}
(see also \cite{tsa60})).
For a general k-essence, with ${\cal L}={\cal L}(\varphi,X)$, 
 equation (\ref{eq2601}) may be written as
\begin{equation}
 c_s^2=\left.\frac{dp}{d\rho}\right|_{s/n}
 =\left.\frac{(\partial p/\partial X)dX +(\partial p/\partial \varphi)
 d\varphi}{(\partial \rho/\partial X)dX +(\partial \rho/\partial \varphi) d\varphi}
 \right|_{s/n} \; \; ,
\label{eq3601}
\end{equation}
where the differentials $dX$ and $d\varphi$
are subject to the constraint $d(s/n)=0$.
Next we show that this constraint implies $d\varphi=0$.

We start from the standard thermodynamical relation
\begin{equation}
 d(\rho V)=
 T d S -p\,dV ,
\label{eq1601}
\end{equation}
where 
the volume 
is, up to a constant factor, given by $V=1/n$.
Equation (\ref{eq1601}) may then be written in the form
\begin{equation}
 dh= T d\left(\frac{s}{n}\right) +\frac{1}{n}dp \; \; ,
\label{eq1603}
\end{equation}
where 
\begin{equation}
 h=\frac{p+\rho}{n}
\label{eq2605}
\end{equation}
is the enthalpy per particle. 
For an  isentropic relativistic flow
one can define a flow potential $\varphi$ such that \cite{land59}
\begin{equation}
 hu_\mu=\varphi_{,\mu} \, .
\label{eq2604}
\end{equation}
 Comparing this with (\ref{eq511})
we find 
\begin{equation}
 h=\sqrt{X}\,. 
\label{eq1602}
\end{equation}
This together with
(\ref{eq4003}) and (\ref{eq4004}) 
 yields in turn
\begin{equation}
 n=2\sqrt{X} {\cal L}_X \, .
\label{eq2603}
\end{equation} 
This expression for the  particle number density is
derived for  an isentropic process. 
In a purely kinetic k-essence with ${\cal L}={\cal L}(X)$,
equation (\ref{eq2603})  follows from
the field equation for $\varphi$
\begin{equation}
 ({\cal L}_X g^{\mu\nu}\varphi_{,\mu})_{;\nu}=0\, ,
\label{eq2702}
\end{equation}
 which implies  conservation of the current 
\begin{equation}
 j_\mu=2{\cal L}_X \varphi_{,\mu} =nu_\mu \, .
\label{eq2602}
\end{equation}
The particle number density $n$ in this expression coincides 
with (\ref{eq2603}).
However, in a general k-essence, with ${\cal L}={\cal L}(\varphi,X)$,
the field equation (\ref{eq07}) does not imply that the current (\ref{eq2602})
is conserved. Nevertheless,  equation (\ref{eq2603}) 
is still a valid expression for
a conserved particle 
number density when the condition $d(s/n)=0$ is imposed.

From (\ref{eq1603}) with $d(s/n)=0$ and
 using (\ref{eq1602}) 
 we obtain
\begin{equation}
 dp=n\,dh=\frac{n}{2\sqrt{X}}dX .
\label{eq2606}
\end{equation}
 Comparing this with the general expression for the total differential of $p$
\begin{equation}
 dp=\frac{\partial p}{\partial X}dX +\frac{\partial p}{\partial \varphi}d\varphi
\label{eq2607}
\end{equation}
   we must have $d\varphi=0$  and
\begin{equation}
 \frac{\partial p}{\partial X}= \frac{n}{2\sqrt{X}}={\cal L}_X \, ,
\label{eq2608}
\end{equation}
as it should be. Hence, we conclude that an isentropic process implies $d\varphi=0$
   and equation (\ref{eq3601}) yields
\begin{equation}
 c_s^2=\left.\frac{\partial p}{\partial\rho}\right|_{\varphi} 
 =\frac{\partial p/\partial X}{\partial \rho/\partial X} \, .
 \label{eq3602}
\end{equation}
\section{Braneworld Connection}
\label{dbrane}
It is useful to view the tachyon condensate from the braneworld  perspective.
Consider a 3+1 brane moving in a 4+1 bulk spacetime with metric
\begin{equation}
ds_{5}^{2} =
g_{(5) MN}dX^MdX^N= 
f(y)^2  g_{\mu \nu} (x)  dx^{\mu}  d x^{\nu} - d y^{2} 
\label{eq4.1}
\end{equation}
where we generalize 
\cite{bil8,jack10,bil39} 
to allow for a warping of the constant $y$  
slices through $f(y)$. The points on the brane are parameterized by 
$X^{\mu} (x^{\mu})$, and $G_{\mu \nu} =
g_{(5) MN}  X_{, \mu}^{M}  X_{, \nu}^{N}$ is the induced metric. 
Taking the Gaussian normal parameterization 
$X^{M} = \left( x^{\mu}, Y(x^{\mu}) \right)$, we have
\begin{equation}
G_{\mu \nu} = f(Y)^2 g_{\mu \nu} (x) - Y_{, \mu}  Y_{, \nu} \, . 
\label{eq4.2}
\end{equation}
The Dirac-Born-Infeld action for the brane is
\begin{equation}
S_{\rm brane} =
-\sigma  \int d^{4}x \sqrt{- \det G_{\mu \nu}} 
  =- \sigma  \int  d^{4}x  \sqrt{-g} \,
  f(Y)^4 
\left[1-\frac{g^{\mu\nu}Y_{,\mu}Y_{,\nu}}{f^{2}(Y)}  \right]^{1/2} , 
\label{eq4.3}
\end{equation}
and with the redefinitions
\begin{equation}
\frac{Y_{, \mu}}{f(Y)} = \varphi_{, \mu}\:, \hspace{1.5cm} \sigma  f(Y)^4 = V(\varphi)
\end{equation}
we obtain
\begin{equation}
{\cal{L}}_{\rm brane} = - V ( \varphi ) 
\sqrt{1 - g^{\mu \nu}  \varphi_{, \mu}  \varphi_{, \nu}} \, .
\label{eq4.5}
\end{equation}
For an unwarped bulk we obtain $f = 1$ and $V = \sigma = \sqrt{A}$, i.e.
 the simple Chaplygin gas.
 In general, $V(\varphi)$
identifies ${\cal{L}}_{\rm brane}$ with the  tachyon model. If the brane also 
couples to, e.g., bulk form fields, there are additional terms that are 
functions of $\varphi$ only.
Thus every tachyon condensate model can be interpreted as a 3+1 brane moving 
in a 4+1 bulk.
Note that this is  true {\it only} for (\ref{eq4.5}) or (\ref{eq049}) but {\it not}
 for  (\ref{eq048}).
The prescription (\ref{eq4.2}) does not take into account the distortion of the bulk metric when the brane is not flat. This, however, can be accounted for using the methods of \cite{kim61}.

Given $V(\varphi)$ the warp factor can be reconstructed via
\begin{equation}
Y - Y_{0} = \int  f(Y ( \varphi))  d \varphi =\sigma^{-1/4}
\int  V(\varphi)^{1/4} d \varphi  .
\label{eq4.6}
\end{equation}
For example, for the power-law potential
\begin{equation}
V( \varphi) = V_{0}  \varphi^{2n} \; \; ,
\end{equation}
we obtain
\begin{equation}
f (Y) = \left( \frac{Y - Y_{0}}{l} \right)^{n/(2+n)}, \hspace{1.5cm}
l = \frac{2}{n+2}\left( \frac{\sigma}{V_{0}}\right)^{1/(2n)} \; \; .
\end{equation}
Using (\ref{eq057}) and writing 
  $\sqrt{G \sigma} = \epsilon/ l$,
  one finds $\epsilon \sim (l H_{0})^{1+n}$
  which is the only fine tuning for $n \neq -1$
  and a small-scale extra dimension.
\section*{Acknowledgments}
We wish to thank Robert Lindebaum for useful discussions.
This research is in part supported by
the Foundation for Fundamental Research
(FFR) grant number  PHY99-1241, the National Research Foundation of South Africa grant number FA2005033 100013, and the Research Committee of the
University of Cape Town.
The work of NB is supported
in part by the Ministry of Science and Technology of the
Republic of Croatia under Contract No. 098-0982930-2864.
 

\end{document}